# Efficient radiational outcoupling of electromagnetic energy from hyperbolic metamaterial resonators

Ildar Yusupov[1,*,+], Dmitry Filonov[2,+], Tatyana Vosheva[2], Viktor Podolsky[3], and Pavel Ginzburg[2,4]

[1]Department of Physics and Engineering, ITMO University, Saint Petersburg, 197101, Russia
[2]Center of Photonics and 2D materials, Moscow Institute of Physics and Technology, 141700 Dolgoprudny, Russia
[3]Department of Physics and Applied Physics, University of Massachusetts Lowell, One University Avenue, Lowell, Massachusetts 01854, USA
[4]School of Electrical Engineering, Tel Aviv University, 69978 Tel Aviv, Israel
[*] ildar.yusupov@metalab.ifmo.ru
[+]these authors contributed equally to this work

## Abstract

Hyperbolic metamaterials were initially proposed in optics to boost radiation efficiencies of quantum emitters. Adopting this concept for antenna design can allow approaching long-standing challenges in radio physics. For example, impedance matching and gain are among the most challenging antenna parameters to achieve in case when a broadband operation is needed. Here we pro-pose and numerically analyse a new compact antenna design, based on hyperbolic metamaterial slab with a patterned layer on top. Energy from a subwavelength loop antenna is shown to be efficiently harvested into bulk modes of the metamaterial over a broad frequency range. Highly localized propagating waves within the medium have a well-resolved spatial-temporal separation owing to the hyperbolic type of effective permeability tensor. This strong interplay between chromatic and modal dispersions enables routing different frequencies into different spatial locations within compact subwavelength geometry. An array of non-overlapping resonant elements is placed on the metamaterial layer and provides a superior matching of localized electromagnetic energy to the free space radiation. As the result, two-order of magnitude improvement in linear gain of the device is predicted. The proposed new architecture can find a use in applications, where multiband or broadband compact devices are required.

## Introduction

   Electromagnetic materials play an important role in tailoring and controlling electromagnetic energy flow. While electrodynamic equations are scalable in respect to an operation frequency, material properties are subject to a considerable chromatic dispersion [1]. As result, architectures of passive electromagnetic devices for optical and centimeter wave ranges are conceptually different. For example, metal components are very attractive for waveguide and antenna design at GHz range because of their almost vanishing losses. For example, copper demonstrates properties of a perfect electric conductor (PEC) within a good approximation. On the other hand, metals are quite lossy at the optical domain and are rarely used in applications where long-range communication is needed. Quite a few efforts have been done in order to control electromagnetic properties of materials by applying different techniques. In this context, the approach of metamaterials allows tailoring permittivities and permeabilities almost on demand by accurate subwavelength structuring of periodic unit cells [2]–[4]. Hyperbolic metamaterials were found to be quite attractive in optical domain, as they suggest achieving relatively high local density of electromagnetic states (LDOS), as it was proposed theoretically [5] and then demonstrated experimentally [6],[7]. The key property, which is responsible for achieving a range of peculiar effects, is an extremely high anisotropy of a susceptibility tensor (ether electric or magnetic). Quite a few realizations of hyperbolic metamaterials have been demonstrated in the optical regime, where free standing metal rods[8]–[10], metal-dielectric layers [11]–[14] or graphene stack [15],[16] are the most widely studied. The physical effect in the beforehand mentioned designs is based on shaping negative permittivity components (e.g. silver, gold, or doped semiconductors), which support surface or localized plasmon resonances [17],[18]. Replicating characteristics of hyperbolic metamaterials from optical to GHz domains is quite challenging since negative susceptibility (either ε or μ) materials are rather scarce in nature. Nevertheless, arrays or loaded wires or split ring resonators were designed and demonstrated sufficient negative values of susceptibility tensors [15],[16],[21]. The drawback of this approach an inherently narrowband operation, which is the direct consequence of the resonance-based methodology. In order to achieve a broadband hyperbolicity at the GHz range, a different design has been proposed. In particular, an array of corrugated wires, supporting spoof plasmon waves, was shown to demonstrate hyperbolic properties in potentially broader spectral window [22]. The possibility to replicate broadband hyperbolic properties into centimeter waves domain suggest considering this material as a component in antenna design.

Material degrees of freedom are rarely considered as tuning parameters in conservative antenna design approaches. In a vast majority of cases, a geometric optimization over a set of shapes is used to achieve demanded characteristics, dictated by an application. Metamaterial approach allows introducing additional design flexibility and assess it versus conventional approaches. For example, high-gain antenna design [23], scattering suppression devices [24],[25], resonators for radiation efficiency enhancement [26], hyperbolic metamaterial-based antennae [27],[28], and several others configurations have been proposed.

An inherent drawback of using electromagnetic materials in antenna applications is the leakage of radiation onto high-index substrates. Similar challenges are faced in solar-cell devices, where light harvesting and photon recycling into semiconductor layers are required. Quite a few different approaches have been developed and include introducing surface roughness, patterned surfaces, arrays of particles (harvesting optical antennas), and others [29]. Light harvesting and extraction from hyperbolic metamaterials also attracted an attention, since standard techniques, described above, can fail. The main reason is high LDOS inside a hyperbolic substrate, which result in highly directional scattering into the bulk [30]. In this case, bulk modes are the preferential scattering channel, and free-space propagating modes are not excited efficiently.

The goal of our investigation on pathways to develop a new generation of antenna devices with metamaterial-based components is to demonstrate a strategy of electromagnetic energy outcoupling from hyperbolic substrates, in some sense bringing the advances of optical science to GHz[31]. Our approach is based on incorporating arrays of resonant scatterers on top of a hyperbolic metamaterial layer (Fig. 1). In this case, an interplay between scattering channels is tailored to maximize the scattering to the free space modes. We have found an optimal balance and have demonstrated more than 100-fold enhancement of extraction efficiency, which can be obtained in several frequency bands, depending on application requirements.

The report is organized as follows: the relevance of hyperbolic metamaterial slot antenna design is discussed first and then followed by a set of optimizations of energy extracting elements. The last part is designated for antenna gain optimization and radiation pattern analysis, which comes before a conclusion.

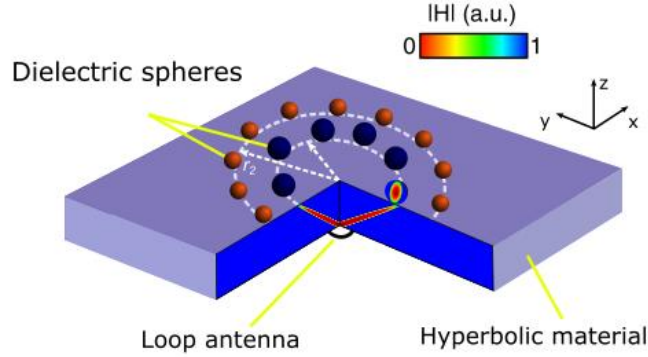

**Figure 1** Structure's layout – hyperbolic metamaterial substrate with circular array of high-index dielectric particles on top. Those arrays are responsible to provide a broadband outcoupling of energy. Each array corresponds to a specific frequency band. The structure is excited by a nonresonant loop antenna, placed underneath the slab.

## Results

An infinite metamaterial slab with an effective diagonal permittivity (Eq. 1) has been considered as a test object.

$$\mu_r = \begin{pmatrix} 1 & 0 & 0 \\ 0 & 1 & 0 \\ 0 & 0 & \mu_{zz}(\omega) \end{pmatrix} \tag{1}$$

Similar susceptibility was demonstrated experimentally in [21], where arrays of near-field coupled split rings resemble an artificial material with negative effective permeability in the vicinity of a resonance. Here, a Drude-type of dispersion for the nontrivial tensor component was assumed (Eq. 2) and it is depicted in Fig. 2(a).

$$\mu_{zz}(\omega) = \mu_\infty - \frac{\omega_p^2}{\omega(\omega - i\omega_c)} \tag{2}$$

where $\mu_\infty = 1$, $\omega_p = 25 \cdot 10^9 \frac{rad}{s}$, $\omega_c = 12 \cdot 10^8 \frac{rad}{s}$

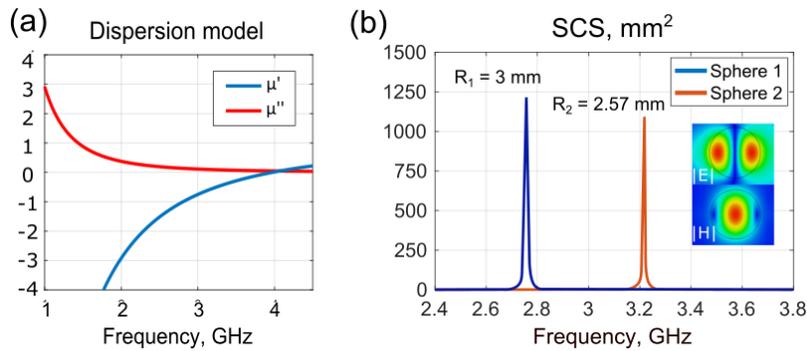

**Figure 2** (a) – dispersion of a nontrivial permittivity tensor component (Eq. 1). Real and imaginary parts are blue and red lines correspondingly. (b) – scattering cross sections of high-index dielectric spheres (radiuses are 3mm and 2.57mm, ε= 330). |E| and |H| field colormaps (arbitrary units) at a magnetic dipole resonance of spheres.

A radiating element (a small 2mm radius loop, connected to a feeding coaxial cable) was brought to a close proximity of the slab. It is worth noting that the radiation efficiency of a deeply subwavelength loop is inherently small. While it can be significantly improved by matching circuits, sufficiently broadband and impedance-matched radiation cannot be obtained without violating Chu-Harrington limit [32] and involving non-Froster matching circuits [33]. Nevertheless, putting an impedance mismatched loop next to a hyperbolic metamaterial results in an efficient excitation of extraordinary modes inside the hyperbolic structures. An unusual (in comparison to free space scenarios) radiation pattern, known as resonance cones[34], can be seen in Fig. 3, is a hallmark of hyperbolic metamaterials (e.g. [35],[36]). A three-dimensional (3D) emission pattern within the slab is a cone with an apex, situated close to the radiating element. The cone hits the upper facet of the slab and re-traced back into the structure. If the inherent ohmic losses of the hyperbolic metamaterial are moderately low, the cone is reflected from the upper surface and additional inverted one is created. Fig. 3 illustrates such multiple reflection scenario, which will be suppressed in the case of moderately high losses. The latter scenario is almost always the case for the optical domain hyperbolic metamaterials (e.g. [8]).

The opening angle of the radiation cone inside the metamaterial has strong chromatic dispersion. This aspect will be the key for demonstrating dual band energy extraction with resonant elements. The angle of the radiation cone can be estimated as a ratio between ordinary and extra-ordinary tensor components:

$$\tan^2\theta(\omega) = \left|\frac{\mu_t(\omega)}{\mu_{zz}(\omega)}\right|,\tag{3}$$

where $\mu_t \equiv \mu_{xx} = \mu_{yy}$ is the tangential component of the permeability tensor. Fig. 3 shows the near-field magnetic field distribution for 2 frequencies, demonstrating the chromatic dispersion of the opening angle, that enables hyper-spectroscopy at optical frequencies[37]. This property will allow obtaining operation at multiple frequency bands and control each channel separately without a need in complex multiplexing circuitry, as it will be shown hereinafter.

After achieving an efficient coupling of energy into the hyperbolic substrate, the next step is to out couple the electromagnetic radiation to the free space[38]. If no additional effort is performed, the upper facet almost completely reflects the electromagnetic energy, leaving an exponentially decaying field in the free space above the surface. In order to solve this extraction problem, an efficient electromagnetic scatterer should perturb this exponentially decaying field and convert it to propagate in the free space. Several parameters should be considered here in order to obtain an efficient design. In particular, a scatterer's position in respect to the surface should be optimized to improve an overall radiation efficiency. For example, if the scatterer is placed too far from the surface, the exponentially-decaying field will not excite it and the reflection back to the substrate will predominate the free space scattering channel. On the other hand, if the scatterer is in the very close proximity to the surface, the re-scattered radiation will leak again into the substrate and not to the free space. Hence, it is reasonable to assume that an engineering tradeoff does exist.

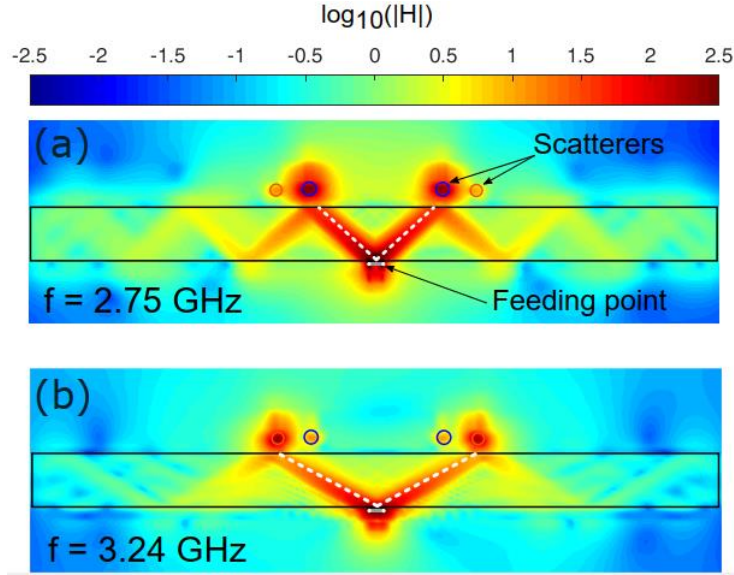

**Figure 3** Electromagnetic energy extraction from hyperbolic metamaterial slab. Logarithmic color maps (arbitrary units) of the magnetic field amplitude (|H|) at two different frequencies of the excitation (a) 2.57 GHz. (b) 3.24 GHZ).

In order to perform the energy extraction optimization, the following setup has been chosen. High-index dielectric spheres were considered as resonating elements. Those structures accommodate moderately high-quality factors along with relatively low internal losses. Those elements can be realized as e.g. silicone nanospheres for optical domain applications (e.g. [39],[40]) and as ceramic resonators for GHz waves (e.g. [41]). In the latter case loss tangents can be as small as $10^{-4}$). High index particles possess a rich family of Mie modes. Here we employ the first magnetic dipolar resonance. Scattering cross-section of a standalone sphere ($R_1 = 3$ mm, $R_2 = 2.57$ mm, $\varepsilon = 330$) appears on Fig. 2(a). and a well-pronounced resonance can be seen. The field distribution of the resonant particle is given at the inset, indicating it's moderately high localization. Dimensions of the particles were calculated in the way they resonate at desired frequencies, chosen for the dual band operation.

The next step is to optimize the distance between the sphere and the metamaterial's surface. This will be a major parameter, affecting the scattering efficiency to the upper half-space. The out coupling process strongly depends on two key parameters, which within a good approximation, can be considered as if they operate sequentially. The first one is the excitation efficiency. A small polarizable sphere is excited by the evanescent field, leaking from the metamaterial substrate (Fig. 4(a)). The dipole moment, which is developed on the particle, is proportional to the sphere's polarizability and the local magnetic field amplitude. Hence, placing the particle far from the boundary will significantly reduce its excitation efficiency. The next step, after the sphere has developed a dipole moment on it, is the scattering. Efficiency of this process can be estimated in separate by placing a small radiating dipole next to the metamaterial boundary. This scenario has been considered analytically in a series of works, related to optical domain applications, e.g. [30],[42]. Main results can be retrieved numerically in a quite straightforward fashion. Fig. 4(b) shows the radiation efficiency (vector Poynting flux through the upper hemi-sphere). Putting a radiating dipole in a close proximity to the hyperbolic substrate results in a strong quenching – the main power flows into the slab, since it supports high LDOS, preferential to scattering. Typical dependency of this quenching is $1/l^3$, where l is the distance to the upper facet. Enlarging the distance causes more efficient radiation to the free space. This factor saturates when there is no near field coupling between the radiating dipole and the slab. The efficiency of the entire process can be estimated as a product of the two factors depicted on panels a and b, indicating the existence of the optimal scatterer's position.

Fig. 4(c) shows the normalized total radiated power (TRP) for the entire scenario. Here 2 circular arrays of equidistant spheres (8 and 12 in inner and outer array, respectively) were considered and the same excitation scheme, used for calculating results on Fig. 3, is employed. The TRP is normalized to the scenario, where the upper facet of the metamaterial is flat and no spheres are present. The ratio demonstrates the overall improvement of energy out coupling. As it was expected the TRP has an optimum for a certain separation distance, which also depends on the operating frequency. This chromatic dispersion is expected since both field decay and particle's polarizability are frequency dependent. In overall, the optimal distance for both cases is between 3-5 mm.

The last stage is the assessment of the dual-band performance of the device. For this purpose we calculated the normalized gain of the structure as a function of frequency. An additional important issue in antenna applications is side lobes, partially affected by the radiation from edges.

Furthermore, front to back ratio also a subject for maximization. A standard technique in antenna design is to enclose a device (apart from its aperture) within a metal shield. Here we place a perfect electric conductor (PEC) on the side and the bottom facets of the structure, leaving the upper plane uncovered. The excitation loop is placed directly within the enclosure. Fig. 5 compares performance of the device with and without arrays of spheres and illustrates the frequency dependence of the normalized antenna gain along with far-field diagrams. Several main features can be clearly seen. The first one is two distinct peaks around the frequencies, which were chosen for the dual band operation. The second aspect is a relatively high gain improvement, 2 orders of magnitude with respect to the unpatterned structure. This result should not be confused with an efficient extraction of energy from high-index layers and does not violate theoretical limits [43]. Furthermore, in order to approach capabilities of a high-performance device, internal material losses should be further reduced. It is also worth noting that the array factor, multiplied by the gain of the single element does not provide a single radiation beam. This results from the fact that a single sphere is predominantly polarized perpendicular to the surface and, consequently, it does not radiate perpendicular to the surface. While this might be an issue in standard antenna applications, the obtained radiation pattern can find a use in radar tracking (e.g. monopulse) applications and several others.

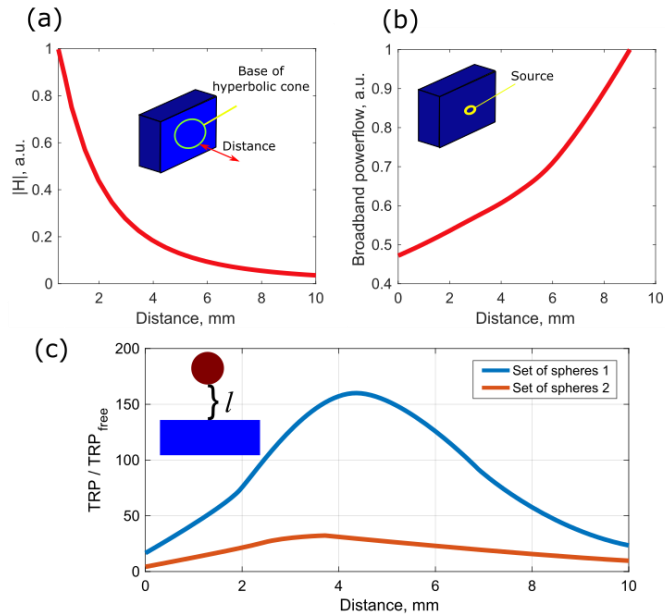

**Figure 4** (a) Normalized magnetic field amplitude, as a function of the distance from the metamaterial surface. The excitation scheme is similar to Fig. 3(a). (b) Power to the upper hemisphere, as a function of a dipolar emitter distance from the metamaterial surface (the radiating dipole is polarized perpendicular to the surface). (c) Total radiated power (TRP) to the upper hemisphere, when the entre scenatio of Fig. 3 is considered (circular array of spheres on top of the metameteril slab). Blue and red curves correspond to the two set of sephres, designed to provide the dual band operation. The presented values are normalized to the total radiaiton efficiency of the device with a flat facet (without any spheres present).

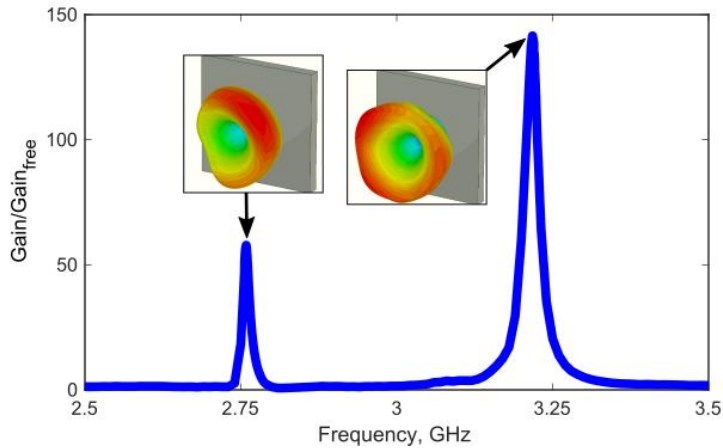

**Figure 5** Normalized gain spectra for the pair of chosen frequencies. The normalization is the ratio of gain, obtained for the device with circular array of spheres and without it. Insets – far-field radiation patterns.

## Conclusion

It was shown that hyperbolic metamaterials can contribute to the endeavor of demonstrating compact broadband antenna devices. While a broadband matching and directivity are rather hard to accommodate together and achieve with standard design approaches, the concept of broadband impedance matching, adopted from the optical domain can break this commonly accepted engineering trade off. Here we demonstrated a design of a dual-band antenna device, which is based on 2 key elements – hyperbolic metamaterial slab for a broadband impedance matching and circular arrays of resonating elements for an efficient out coupling. The optimized structure show more than 2 orders of magnitude improvement of the antenna gain (linear scale) and paves a way for further improvement of other antenna characteristics. The key step for demonstrating the proposed concept in antenna applications is a metamaterial realization with moderately low internal losses and miniature unit cell. The later factor is essential to ensure applicability of the effective medium model. Furthermore, the unit cell granularity should be finer than the size of the resonating elements, used for the radiation outcoupling. For example, realizations, demonstrated in [21] can be further miniaturized with meandering techniques, enhanced by lumped elements, plugged in relevant parts of resonating elements.

## Competing interests

The authors declare no competing interests.

## Acknowledgements

The research was supported in part by Binational Science Foundation (project 2016059); PAZY Foundation and the Russian Science Foundation (Project 20-19-00480)

## Author contributions

All authors reviewed the manuscript.